\documentclass[aps,prb,twocolumn,superscriptaddress,showpacs]{revtex4-2}

\usepackage{graphicx}                             
\graphicspath{ {./_figures/} }   

\usepackage{amsmath}
\usepackage{amssymb}
\usepackage{bm}

 \usepackage{color}                                     
\usepackage{hyphenat}

\begin{document}


\title{X-ray detected ferromagnetic resonance techniques\\ for the study of magnetization dynamics}
 
\author{Gerrit {van der Laan}}
\affiliation{Diamond Light Source, Harwell Science and Innovation Campus, Didcot, OX11 0DE, United Kingdom}

\author{Thorsten Hesjedal}
 \affiliation{Department of Physics, Clarendon Laboratory, University of Oxford, Oxford, OX1 3PU, United Kingdom}


\begin{abstract}
Element-specific spectroscopies using synchrotron-radiation  can provide unique insights into materials properties.
The recently developed technique of X-ray detected ferromagnetic resonance (XFMR)   allows studying the magnetization dynamics of magnetic spin structures. Magnetic sensitivity in XFMR is obtained from the X-ray magnetic circular dichroism (XMCD) effect, where the phase of the magnetization precession of each magnetic layer with respect to the exciting radio frequency is obtained using stroboscopic probing of the spin precession.  Measurement of both  amplitude and phase response in the magnetic layers as a function of bias field can give a clear signature of spin-transfer torque (STT) coupling between ferromagnetic layers due to spin pumping.
Over the last few years, there have been new developments utilizing X-ray scattering techniques to reveal the precessional magnetization dynamics of ordered spin structures in the GHz frequency range.
The techniques of diffraction and reflectometry ferromagnetic resonance (DFMR and RFMR) provide novel ways for the probing of the dynamics   of chiral and multilayered magnetic materials, thereby opening up new pathways for the development of high-density and low-energy consumption data processing solutions.
\end{abstract}

\date{\today}
\maketitle




\section{Introduction}
 
Magnetization dynamics is at the heart of high frequency magnetic nanoscale devices based on spin waves, spin pumping, and spin-torque oscillators in the GHz frequency range. 
Traditionally,  ferromagnetic resonance (FMR) has been a work horse technique to determine the fundamental parameters for magnetic resonance and relaxation in thin films.  The recent growing complexity of many modern  magnetic materials and devices requires the development of  advanced measurement techniques that more directly reveal the microscopic origin of the dynamical magnetic interactions that are at play.

The novel techniques of  X-ray detected FMR (XFMR) enables studying the magnetization dynamics of individual layers, where element\hyp{}specific magnetic contrast  is obtained using the X-ray magnetic circular dichroism (XMCD) effect \cite{Vanderlaan2017}. Not only can the FMR signal  be monitored in X-ray absorption, it can also be done in X-ray diffraction and reflectivity, using techniques   termed as DFMR and RFMR, respectively \cite{Burn2021}.
In these X-ray measurements,   time-resolved FMR gives both the amplitude and phase of the spin precession for the different chemical elements, and hence different layers,  in the sample.
The challenge of such measurements lays in the fact that the  precession cone angle is small ($<$$1^\circ$) and  that the precession frequency is on the order of GHz. 
The solution is to use lock-in techniques and to detect the phase of the precession stroboscopically by using the time structure of the X-ray pulses from the synchrotron ($\sim$500 MHz, i.e., corresponding to a period  between the pulses of 2 ns).  The radio frequency (RF) field applied to drive the spin precession is synchronized with the X-ray pulses using the clock of the synchrotron. Therefore, each X-ray pulse measures the magnetization cone at precisely the same point in the precession cycle.
Hence, XFMR combines the techniques of FMR and XMCD. Thus, the spin precession along the bias field  is pumped by the  RF field   to generate the magnetic resonance (i.e., FMR), whose amplitude and phase is probed by the synchronized  X-ray pulses using the XMCD effect. The time dependence is recorded using a delay line to vary the phase of the RF signal with respect to the X-ray pulses.

During the last few years, many XFMR studies either in time-averaged or time-resolved mode have been reported
\cite{bailey2004_PRB,Arena2006_PRB,Guan2006,arena2007_JAP,Guan2007,Arena2009_SRI,
bailey2013_NC,warnicke2013,warnicke2015, 
Goulon2005,Goulon2006,goulon2007_JSR,goulon2009_JMMM,Goulon2010,goulon2011_JMS,goulon2014_JMMM,
boero2005_APL,Boero2008_NJP,Boero2009_PRB,Boero2009_RSI,
Marcham2011,Marcham2013APL,Marcham2013PRB, li2016_PRL,
baker2015_SR,baker2016_PRL,baker2016_SR,figueroa2016_JMMM,stenning2015,
klaer2011,martin2008_JAP,martin2009_JAP,salikhov2011_APL,salikhov_PRB2012, chou2006_JAP,
ollefs2015_JAP,bonetti2015_RSI,Durrant2017, Baker2019,Dabrowski2020-PRL, Klewe2020,
Dabrowski2020-AMI, Gladczuk2021-PRB, Dabrowski2021, Gladczuk2021-PSS, Yang2022,Klewe2022, Lim2022,
Emori2020, Pogoryelov2020
}.
The first element-specific and time\hyp{}dependent measurement of the magnetization dynamics using pump-probe  XMCD was reported  by  Bailey {\it{et al.}}~\cite{bailey2004_PRB} on a permalloy (Py = Ni$_{80}$Fe$_{20}$) thin film, where  the moments on the Ni and Fe sites were found to precess together at all frequencies, and by Arena {\it{et al.}}~\cite{Arena2006_PRB} on a Py/Cu/CoZr trilayer, where at resonance, a weak ferromagnetic coupling  was found in the phase and amplitude response of individual layers across resonance.

The amplitude and phase response  of the magnetic probe  layer measured by XFMR provides a  signature for either
static exchange interaction in strongly exchange-coupled bilayers or spin-transfer torque (STT) coupling due to spin pumping.  
 Marcham  {\it{et al.}}~\cite{Marcham2013PRB} first evidenced STT in a CoFe/Cu/Py spin valve  using  XFMR where the field dependence of the fixed layer phase showed a clear signature of STT due to spin pumping. 
Using XFMR, Baker  {\it{et al.}}~\cite{baker2016_PRL} reported a strong anisotropy of the spin pumping, providing new opportunities for device applications.

Previously, time-resolved XFMR has  been reviewed in great detail in Ref.~\cite{Vanderlaan2017}.
Here, we  present a timely update,  especially emphasizing the newly developed  time-resolved FMR techniques in X-ray reflectivity and diffraction.
 
The outline of the remainder of this paper is as follows. 
Sec.~\ref{sec:theory} gives a brief theoretical background of magnetization dynamics and STT.
Sec.~\ref{sec:experimental} describes the experimental setup, conditions, and considerations for the various XFMR techniques.
Sec.~\ref{sec:examples} highlights several recent examples of XFMR, DFMR, and RFMR experiments and mentions their scientific impact.
Finally, conclusions are drawn in Sec.~\ref{sec:Conclusions}.

\section{Background on FMR and STT} \label{sec:theory}

\subsection{Ferromagnetic resonance (FMR)}

Before presenting the experimental details and showcasing several recent examples, we will briefly introduce some relevant background material.

FMR arises when the energy levels of a quantized system of electronic moments are Zeeman split by a uniform magnetic field and the system absorbs energy from an oscillating magnetic field \cite{Morrish1965}. A resonance occurs when the transverse AC field is applied at the Larmor frequency corresponding to the energy difference between the magnetic levels, i.e., $\hbar \omega = \Delta E$.   
 The spin precession in a single-domain magnetic material can be described with the equation of motion, the so-called Landau-Lifshitz-Gilbert (LLG) equation,
 \begin{equation}
\mathbf{ \dot{m} }  = -    \gamma  \mathbf{m} \times \mathbf{H}_{\mathrm{eff}}+ \alpha    ( \mathbf{m} \times \mathbf{\dot{m}} ) ,
\label{eq:LLG2}
\end{equation}
where the effective field $\mathbf{H}_{\mathrm{eff}}  = - \partial F (\mathbf{M} )/\partial \mathbf{M}$ is  obtained by minimization of the free energy $F$ with respect to the magnetization  ${\bf{M}}$.
The free  energy contains terms such as the  exchange, Dzyaloshinskii-Moriya, demagnetization, magnetocrystalline
anisotropy, magnetostatic, external Zeeman field, and elastic energy.
Further,  $\mathbf{ \dot{m} } = \delta \mathbf{ m } / \delta t$; the reduced magnetization is $\mathbf{m} = \mathbf{M}/ M_s $,  
where $M_s = | \mathbf{m} | $ is the saturation magnetization; 
and $\gamma = g \mu_{\mathrm{B}} / \hbar$ is the gyromagnetic ratio, where $\mu_{\mathrm{B}} $ is the Bohr magneton and $g$ is the Land\'e (spectroscopic splitting) g-factor.
The  dimensionless damping parameter  $\alpha \ll 1$ (typically  10$^{-3}$--10$^{-2}$ for $3d$ metals) determines the width of the resonance absorption peak. 

The first right-hand term in Eq.~(\ref{eq:LLG2}) corresponds to the torque   due to the effective field $\mathbf{H}_{\mathrm{eff}}$. In a classical picture, $\tau = d \mathbf{S}/dt$  equates to the time change in angular momentum $\mathbf{S}$, which leads to the spin precession. The second  right-hand term corresponds to the  damping,
which can also be written in the form of the Gilbert damping term $ - \alpha    \gamma ( \mathbf{m} \times \mathbf{m} \times {\mathbf{H}}_{\mathrm{eff}})$. Both torque and damping are vectorially sketched in Fig.~\ref{fig:LL}(a).
Without  external RF excitation, the magnetization would relax to the steady state given by Brown's equation, $\mathbf{m} \times {\mathbf{H}_\mathrm{eff}} = 0$.

Linearization of the LLG equation gives the relation between the frequency $ \nu_0$ (or circular frequency $ \omega_0$) and field, which in the form of the Kittel equation is written as \cite{Vanderlaan2017}
\begin{equation}
 2 \pi   \nu_0 \equiv  \omega_0 
 =  \gamma  \sqrt{ {H_\mathrm{eff}} {B_\mathrm{eff}} }
=  \gamma  \sqrt{ {H_\mathrm{eff}} (M_s +{H_\mathrm{eff}} )} .
\label{eq:Kitteleq}
\end{equation}
 
\subsection{Spin-transfer torque (STT)}
\label{sec:SST}

The layer selectivity of  XFMR  makes this technique a unique probe to investigate STT and related spin currents in multi-layered spin valves \cite{Vanderlaan2017}.
STT is the effect in which the spin direction in a magnetic layer  can be modified using a spin-polarized current \cite{slonczewski1996,berger1996}.
%
%

Spin pumping occurs when the precessing magnetization vector generated  by FMR  in a ferromagnetic (FM) layer emits a pure spin current into an adjacent normal metal (NM) layer \cite{tserkovnyak_spinPumpingReview}.
Traditionally, spin currents have been probed using indirect measurements.  For instance, in the metals through which they flow they can create an electrical voltage drop perpendicular to the spin current direction, or a torque that bends the magnetization direction. However, such indirect measurements are often ambiguous because they are also influenced by other factors, such as magnetic proximity effects at the  interface.

\begin{figure}[t]				               
\centering              
\includegraphics[trim =0mm 0mm 0mm 0mm,clip, width=85mm, angle=0]{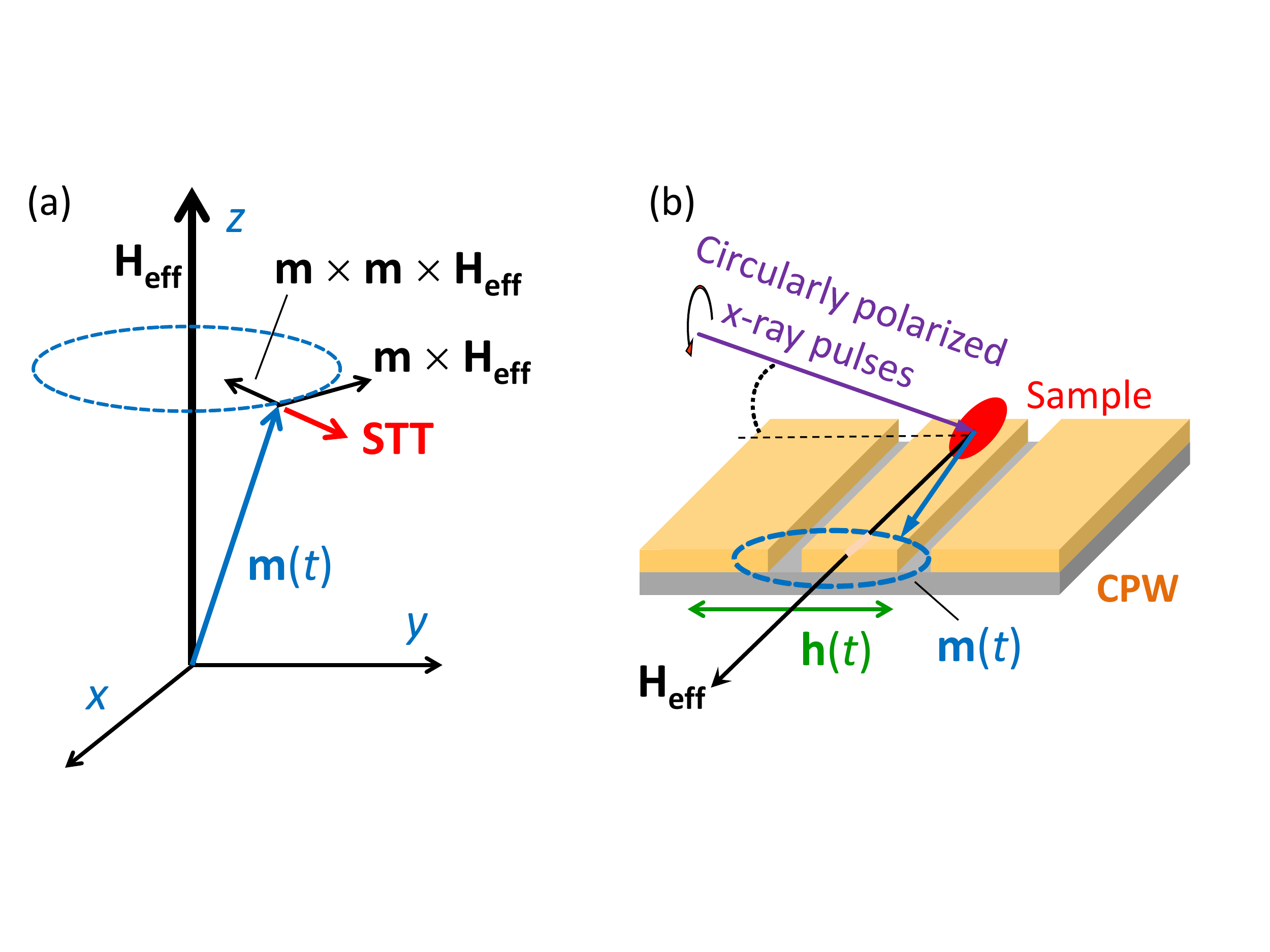}
\caption{(a) Precession, damping, and spin transfer torque (STT) in FMR.
The precession $\mathbf{m} \times \mathbf{H}_{\mathrm{eff}}$ around  the effective field $\mathbf{H}_{\mathrm{eff}}$ is damped by   the Gilbert term 
$ \mathbf{m} \times   \mathbf{m} \times   \mathbf{H}_{\mathrm{eff}}$. 
The spin-transfer  torque  is  parallel (antiparallel) to  the Gilbert damping term,
 and can enhance (oppose) the latter depending on the direction of the spin current.
 (b) Schematics of the sample geometry for XFMR.
 The sample (red disk) is mounted on the signal line (the central strip) of a coplanar waveguide (CPW). The magnetization ${\mathbf{m}}(t)$ precesses  about   ${\mathbf{H}}_{\mathrm{eff}}$, driven by the in-plane continuous RF field  ${\mathbf{h}}(t)$  in the CPW. 
 The cone angle of precession is exaggerated for clarity; its typical magnitude is $\sim$1$^\circ$. 
 Circularly  polarized X-ray pulses from the synchrotron impinge at an grazing incidence angle on the sample in transverse geometry in order to enable  stroboscopic detection of the oscillatory component of ${\mathbf{m}}(t)$ at variable phase delays.
 }
\label{fig:LL}
 \end{figure}

STT  gives   an extra term in the LLG equation, which is (anti)-parallel to the (anti)-damping 
(see  Fig.~\ref{fig:LL}(a)).
According to Slonczewski  \cite{slonczewski1996}, the adiabatic torque  is 
$ \mathbf \tau_{\mathrm{s}} =  \alpha^{\mathrm{s}}  \,  \mathbf{m}   \times \mathbf{\dot{m}}$,
where $\alpha^{\mathrm{s}}$ is the STT damping.
 %
The spin current pumped across a FM/NM interface due to precession is \cite{tserkovnyak_spinPumpingReview}
\begin{equation}
\mathbf{I}_{\mathrm{s}} = \frac{\hbar } { 4 \pi} g_{\mathrm{eff} }^{\uparrow \downarrow} \  \mathbf{m}   \times \mathbf{\dot{m}} \,\,,
\end{equation}
where $g_{\mathrm{eff} }^{\uparrow \downarrow} $ is the effective spin-mixing conductance.
The  spin pumping depends critically on  the FM/NM interface (the material-dependent  $g^{\uparrow \downarrow}_{\mathrm{eff}}$) and  the  spin diffusion length in the NM.

For two FM layers $i$ and $j$ with different resonance frequencies and  coupled  by both  spin pumping (dynamic exchange coupling) and   static exchange coupling, the coupled LLG  equations are 
\begin{align}
\mathbf{\dot{m}}_i   = &   - \gamma \mathbf{m}_i \times \mathbf{H}_{\mathrm{eff},i} 
+ \alpha_i^0 \mathbf{m}_i \times \mathbf{\dot{m}}_i   \nonumber   \\
& +  \alpha^{\mathrm{s}}_i  \mathbf{m}_i \times ( \mathbf{\dot{m}}_i  - \mathbf{\dot{m}}_j ) 
 + A_{\mathrm{ex}} \mathbf{m}_j \cdot \mathbf{ m}_i   \  ,
\label{eq:LLGS}
\end{align}
and equivalently when exchanging $i  \leftrightarrow  j$,
where $\mathbf{ m}_i $ is  the magnetization direction,  $\mathbf{H}_{\mathrm{eff},i}$    the effective field,
 $\alpha^0_i$ the Gilbert damping, and  $ \alpha^{\mathrm{s}}_i $  the STT damping  in layer $i$. The spin pumping induced coupling is determined by $ \alpha^{\mathrm{s}}_i $ and the static exchange coupling by $A_{\mathrm{ex}}$.

\section{Experimental}
\label{sec:experimental}


 XFMR provides an element-specific  and
time-resolved measurement of  the precessional dynamics of each FM layer on  a ps time scale, where
the  spin precession induced by a driving RF signal is detected using the XMCD effect \cite{GVDLAFG2014}.
However, before performing the XFMR experiment, the static magnetization of the samples has to be  precharacterized with 
 standard techniques, such as superconducting quantum interference device (SQUID) magnetometry to measure the hysteresis loops along the easy  and hard direction, followed by standard FMR measurements.

\subsection{VNA-FMR}

Vector network analyzers (VNA)-FMR measurements are used to characterize the magnetic resonances in order to judge whether these are suitable and intense enough for  the XFMR   measurements at the synchrotron. VNA-FMR is a broadband FMR technique, where the sample is mounted onto a coplanar waveguide (CPW) and driven by an external RF field, while under a static magnetic bias field. 
Measurement of the $S$-parameters of the sample results in a frequency-field map, where the resonances  appear in the form of  Kittel curves  (Eq.~(\ref{eq:Kitteleq})). 
The angular dependence of the resonances gives information about the magnetic anisotropy \cite{Morrish1965}. 
It allows us to chose the best azimuthal angle of the applied field with respect to the crystallographic axes to separate the magnetic resonances at the optimal distance for detecting STT \cite{baker2016_PRL}.
Hence, at a given RF frequency, this gives us the corresponding field values of the resonances in the XFMR experiments.
Conventional FMR will normally probe the whole thickness of a thin film since the skin depth of, e.g.,  metallic iron at 10 GHz, is on the order a micron.

\subsection{XMCD}

At the synchrotron, first  the static XMCD is measured by sweeping the photon energy across the absorption edge  of the magnetic elements. This allows us  to select the fixed photon energies  suitable for XFMR. 
The static XMCD is obtained from the difference between  two X-ray
absorption spectra recorded with the helicity vector of the circularly
polarized X-rays parallel and antiparallel, respectively, to the applied
magnetic field \cite{GVDL2013}. The XMCD signal is proportional to
the projection of the helicity vector, which is along the incident beam
direction ${\hat{\mathbf{k}}}$, onto the sample magnetization $
{\mathbf{M}}$, hence $I_{\mathrm{XMCD}} \propto {\hat{\mathbf{k}}}
\cdot {\mathbf{M}} $.

The XMCD  at the soft X-ray absorption edges, such as the  Fe, Co, and Ni  $L_{2,3}$,   is very strong  \cite{GVDL1991}, which helps to compensate for the small changes in magnetization direction due to the limited cone angle ($<$$1^\circ$) of the precession in  XFMR.
 The X-ray  penetration length, which limits the sampling depth, is in the nm range, e.g., for pure Fe it is  $\sim$20 nm at   the Fe $L_3$   maximum at $\sim$707 eV \cite{GVDLAFG2014}. By tuning the photon energy away from the absorption maximum the penetration length can be increased ($\sim$600 nm below the edge at 700 eV).
Note that the length scale of the probing depth  is well matched to the thickness of the magnetic layers in spin valves.
The typical lateral spot size of the X-ray beam  on the sample is 200 $\times$ 20 $\mu$m$^2$, again well suited for small devices.

\subsection{Time-resolved measurements}

The measurement of the projected magnetic moment in XFMR  does not require to take the difference  between opposite circular polarizations as done in XMCD. Instead, a change in the projection of the magnetization precession is measured  using a fixed circular polarization. 

 XFMR can  be measured in two distinctly different geometries, namely ({\it{i}})
time-averaged in longitudinal geometry
\cite{Goulon2005,Boero2008_NJP} or ({\it{ii}}) time-resolved in transverse
geometry \cite{Arena2009_SRI,Marcham2013PRB,baker2016_PRL}. 
 In longitudinal geometry, a shortening of
the magnetization vector along the $z$-axis (parallel to the X-ray beam direction) leads
to a difference $\Delta M_z$ = $M_s (1-\cos \theta) \approx
\frac{1}{2} M_s \theta ^2$, where $\theta$ is the small cone angle of the magnetization precession. The time-averaged XFMR requires no
synchronization with the synchrotron clock, therefore it can be done
at an arbitrary frequency, but it needs a larger RF power which can lead to nonlinear effects.

Only measurements in  transverse geometry  give access to the precessional phase.
 This geometry is depicted in Fig.~\ref{fig:LL}(b).
The transverse component of the magnetization precession will give a   sinusoidal variation on top of the static X-ray absorption signal.
With the  incident X-ray beam 
perpendicular to the bias field, the oscillating component
of the magnetization precession is measured with a magnitude $|M_y|$ = $M_s
\sin \theta \approx M_s \theta$. Thus, for a typical cone angle of $\theta \approx 1^\circ$, the transverse geometry gives a  signal  that is larger  by a factor  of $\sim$200 compared to the longitudinal geometry. 
 Due to the shape anisotropy of the film, the precession  is
strongly elliptical, often with a larger in-plane amplitude.  
 This favors a measurement geometry with the X-rays at  grazing incidence. 
A good compromise is an X-ray incidence angle of $\sim$35$^\circ$ with respect to the plane of the sample, which ensures that the signal is sensitive to the larger in-plane component of the magnetization precession. 

Using a vector magnet system, such as the portable octupole magnet  system (POMS) at Diamond \cite{GVDLAFG2014}, where the field can be
applied in any direction,   permits a simple
change of the field from ({\it{i}}) parallel to the photon
direction,  as needed for static XMCD scans, to ({\it{ii}})
orthogonal to both X-ray beam  and  RF field direction,
as required for time-resolved XFMR.

The detection of the X-ray absorption can be done by either    X-ray transmission \cite{stenning2015,baker2015_SR}, fluorescence yield \cite{Marcham2011}, or X-ray scattering or reflectivity \cite{salikhov2011_APL,salikhov_PRB2012,Burn2020-DFMR,Burn2020-RFMR,Burn2021}. However,  RF  plays havoc with total-electron yield.
In the case of transmission, the incident X-rays impinge on the sample through a hole in the signal line of the CPW. 
After passing through the sample, the  transmitted X-rays are detected with X-ray excited optical luminescence (XEOL) emerging from the MgO or sapphire (Al$_2$O$_3$) substrate  using a photodiode
placed behind the sample. Note that not all substrates, such as non-transparent ones like Si,  are  suitable for XEOL detection \cite{vaz2013}.

\begin{figure}[t]				               
\centering             
\includegraphics[trim =0mm 0mm 0mm 0mm,clip, width=85mm, angle=0]{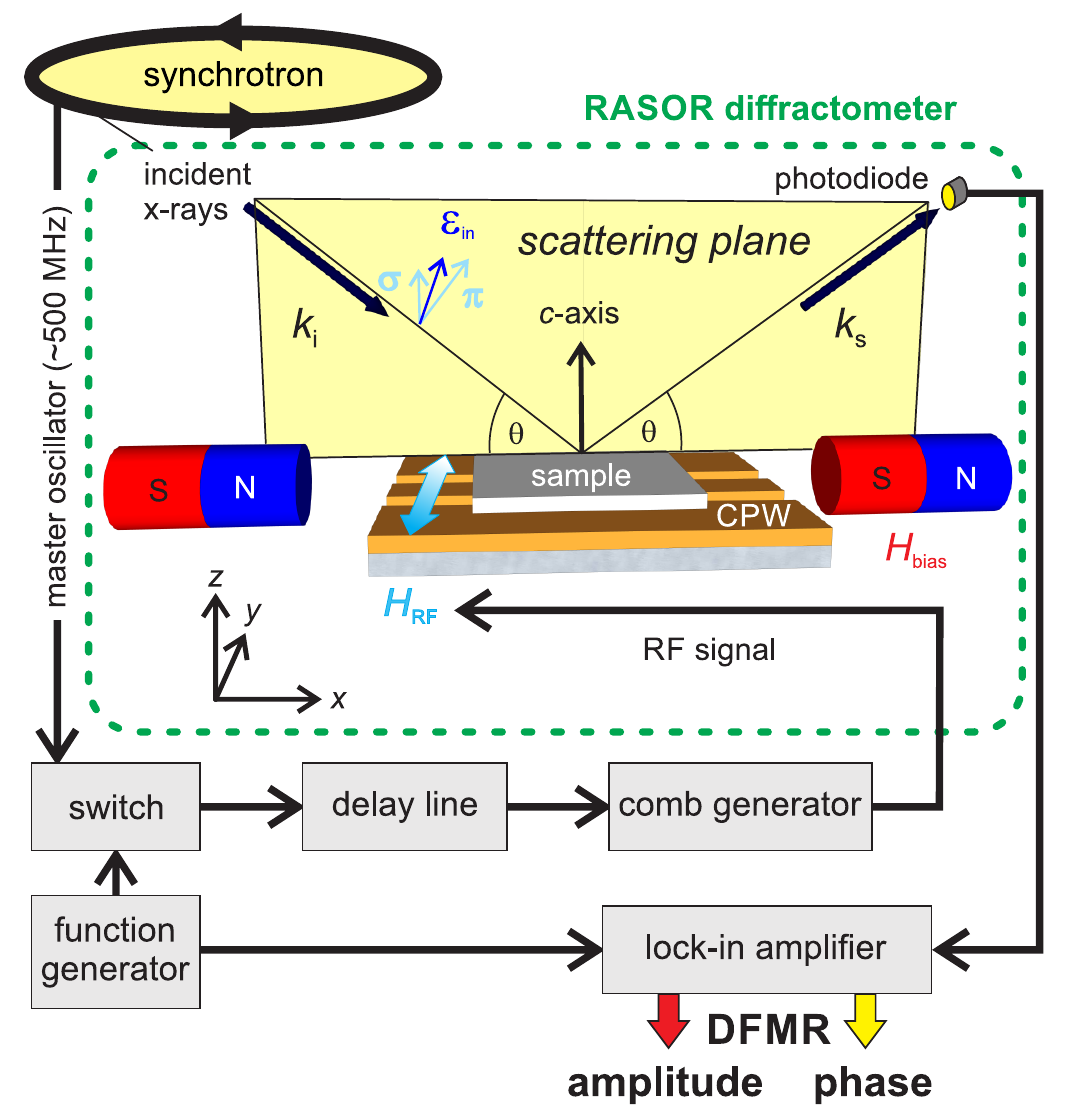}
\caption{
Schematic of the setup for DFMR measurements in the RASOR
diffractometer at the Diamond Light Source. The sample
is placed on the CPW, which is mounted on the cold finger inside
the diffractometer. Incident circularly or linearly polarized X-rays are scattered off the sample 
and detected via a photodiode in a $\vartheta$-$2 \vartheta$ geometry. 
A variable magnetic field is applied in the scattering
plane via a pair of permanent magnets whose distance can be controlled externally.
An RF signal is fed to the CPW to drive the ferromagnetic
resonance in the magnetic sample. 
As the synchrotron gives X-ray pulses at a frequency of $\sim$500 MHz,
a comb generator is used to produce higher harmonics, which are
selected and fed to the CPW. To probe the time dependence of the
scattered X-ray intensity, a tunable delay line is used, which shifts
the phase between the pump (the RF signal) and the probe (the pulsed X-rays).
(Adapted from Ref.~\cite{Burn2022}).
 }
\label{fig:setup}
 \end{figure}

Time resolution is established by  using the periodic X-ray pulses from the synchrotron (normally operating  in multibunch or hybrid mode).
To enable stroboscopic probing, the RF driving field is taken as a harmonic of the
X-ray pulse frequency, hence the resonance is driven at multiples of
the master oscillator clock of the storage ring. 
These  harmonics are generated using an RF comb generator (Atlantic
Microwave) driven by the master oscillator clock, which has a frequency of
499.65~MHz (at the DLS,  ALS, and BESSY synchrotron). This corresponds to $\sim$2~ns
intervals between consecutive X-ray pulses, which have   a pulse width of $\sim$35 ps (at DLS  and BESSY) or $\sim$70 ps (at ALS). The desired frequency is
selected using filters and amplifiers to drive a narrow band, high
power (25--30~dBm) RF field to the CPW. A programmable delay line
(Colby Instruments) enables phase shifting of the RF oscillation
with respect to the X-ray pulses with a step resolution of
$\sim$0.5~ps.
Depending on the specific technique either the transmitted, diffracted, or reflected X-rays are measured using a photodiode.
Fig.~\ref{fig:setup} show a schematic representation of the setup for DFMR; for XFMR and RFMR the electronics is very similar.
The  signal is obtained using a lock-in amplifier (LIA) by switching the signal at a given audio frequency.  
There are two usual modulation modes.
 In amplitude modulation the LIA measures the difference between signals obtained with the RF signal on and off. 
In  180$^\circ$-phase modulation,  the LIA measures the difference between signals obtained with the RF  of opposite phase.

\subsection{XFMR}

In order to record  the time-resolved XAS signal with circular polarization at  fixed photon energy,
the RF frequency is locked to a multiple of the synchrotron  clock.  Then at fixed angles and for given  temperature, this leaves   two free scanning parameters, namely the magnetic bias field strength and the delay time between  X-ray pulses and  RF field.

{\it{Magnetic field scans}} record the signal by sweeping the  bias field at a constant delay time. The signal contains both real and imaginary parts of the magnetic susceptibility, whose relative contributions strongly change across resonance. By  measuring two field scans, which differ by  90$^\circ$  in phase (obtained using the corresponding time delays), and fitting these scans simultaneously using the Kramers-Kronig relation,  gives a good
apprehension of the field dependence of the resonances  \cite{Durrant2017}. 

{\it{Delay scans}} record the signal for each of the magnetic layers at  constant bias field by sweeping the delay time.  
As an example, Fig.~\ref{fig:Gladczuk}(a) shows a series delay scans over two periods of the phase taken at different bias fields (40--200 mT) across the Co   resonance in a magnetic tunnel junction (MTJ), in more detail discussed in  Sec.~\ref{sec:Gladczuk}. 
The solid lines represent sinusoidal fits to the experimental data (dots), from which the amplitude and relative phase  of the  magnetization precession can be extracted. 
A sinusoidal function of the form
\begin{equation}
S(t) = X \sin(2 \pi \nu t) + Y \cos(2 \pi \nu t) ,
\label{eq:sinusoidal}
\end{equation}
is fitted to the delay scan, where $t$ is the time delay and $\nu$
the frequency of the RF. This procedure
is repeated for various field strengths and directions. By
extracting the coefficients $X$ and $Y$ in Eq.~(\ref{eq:sinusoidal}) from the delay
scans, the amplitude $A$ and phase $\psi$ of the oscillations can be
determined using the relationships
\begin{equation}
A =\sqrt{X^2+Y^2}, \  \  \  \  \psi = 2 \arctan \left( \frac{Y}{A+X} \right) .
\label{eq:A-psi}
\end{equation}

{\it{XFMR  precessional plots}} are assembled by combining the amplitudes and phases extracted from the delay scans measured over a range of bias fields. This gives the field dependence of the amplitude and phase for each element (e.g., for Co and Ni in Fig.~\ref{fig:Gladczuk}(b)), from which the type of coupling between  layers can be assessed.
By normalizing the XFMR signal to the static XMCD, the amplitude of the signal can be obtained per atom for each chemical element in the sample. This enables a quantitative decomposition of the resonance features \cite{stenning2015}. 
 
The static coupling (i.e., exchange interaction) and dynamic coupling  (i.e., spin pumping)  give a very different  XFMR response, as can be understood from Eq.~(\ref{eq:LLGS}).
Consider a pump layer FM1 that is free to rotate, and a probe layer FM2 that is pinned.
Using XFMR  at a fixed frequency, we scan the field across the entire resonance.
At resonance, FM1 will show a symmetric  peak for the amplitude, while the phase is 90$^\circ$ delayed with respect to the RF driving field. 
Across the entire resonance, the phase will change by 180$^\circ$. 
To investigate the type of coupling between both layers we measure the XFMR response of FM2  at the resonance condition of FM1.

For static exchange coupling,  $E = -A_{\mathrm{ex}}\mathbf{m}_1  \cdot   \mathbf{m}_2$, so that 
$\mathbf{H}_{\mathrm{eff},2} \propto \mathbf{m}_1$. This means that the effective field in the second layer is aligned along the magnetization of the first layer.
Then the field dependent precession of  FM2 will show a  dispersive  (bipolar) peak in the amplitude and a symmetric (unipolar) peak in the phase.
 
On the other hand, for dynamic exchange coupling
$\mathbf{H}_{\mathrm{eff},2} \propto   \mathbf{\dot{m}} _2$ = $- \mathrm{i} \omega \mathbf{m}_1$.
The magnetic field is imaginary,  resulting in  a 90$^\circ$ phase change. 
In this case,  the  field dependent precession of FM2 will show  a unipolar  peak in the amplitude and   bipolar peak in the phase.
 
This behavior means that XFMR can distinguish between static and dynamic coupling by their amplitude and phase signature in the precessional plot, and thus determine the  relative contribution of these couplings. This has  previously been utilized for, e.g., 
 exchange coupled layers \cite{Arena2006_PRB,stenning2015}, 
  spin values \cite{Marcham2013PRB}, 
 MgO  magnetic tunnel junctions \cite{baker2016_SR,Gladczuk2021-PRB},
 topological insulators  \cite{baker2015_SR,figueroa2016_JMMM,Baker2019},
 spin valve with $\delta$-layer  \cite{li2016_PRL},
 Heusler alloys  \cite{Durrant2017},
NiO antiferromagnetic interlayer \cite{Dabrowski2020-PRL},
exchange springs \cite{Dabrowski2020-AMI}, and
$\alpha$-Sn thin films \cite{Gladczuk2021-PSS}.

 \subsection{DFMR and RFMR}
 
 DFMR and RFMR measurements have been performed in the RASOR soft X-ray diffractometer on beamline I10 at the Diamond Light Source \cite{GVDLAFG2014} (see setup in Fig.~\ref{fig:setup}).  Incident X-rays with wavevector $\mathbf{k}_i$ illuminate the sample, while the scattered beam ($\mathbf{k}_s$) is detected using a photodiode. The scattering geometry  is configured to probe the sample at certain diffraction or specular reflectivity conditions. The sample  in the diffractometer  is mounted on a CPW  that is connected to a liquid He cryostat arm which can reach temperatures down to 12 K. A bias magnetic field is applied by two permanent magnets, which can be positioned to vary both the field strength up to 200 mT and the orientation within the scattering plane.
Perpendicular to the bias field, a transverse RF  field around the central conductor of the CPW is generated, which excites the magnetization dynamics in the system.  In contrast to conventional XFMR measurements, where the sample is mounted flip-chip onto the CPW, in the scattering geometry the sample is mounted face up to allow for the X-ray beam to probe its surface. To ensure good coupling between the CPW and the probed top surface, the sample must either be thinned, or in the case of multilayers, grown on a thin substrate of the order of 100 $\mu$m.

In DFMR, where the detector is aligned to a Bragg peak or  magnetic scattering peak,  the stroboscopic signal  is used to measure delay scans for different linear or circular polarization of the incident X-rays, to give information about the periodic spin structure.

In RFMR where the photo diode detector accepts the reflected beam  the stroboscopic signal  is used to measure delay scans for different values of the scattering vector $Q_z$, to obtain depth information. 
An advantage of RFMR over DFMR is that it can be done on thin films and multilayers, and no single crystals are needed.

\section{X-ray based FMR examples}
\label{sec:examples}

\subsection{XFMR of spin-current mediated exchange coupling in MgO-based MTJs}
  \label{sec:Gladczuk}
 
 Magnetic tunnel junctions composed of ferromagnetic layers which are mutually interacting through a nonmagnetic spacer layer are at the core of magnetic sensor and memory devices. 
 G{\l}adczuk {\it{et al.}}\ \cite{Gladczuk2021-PRB}  used layer-resolved XFMR to investigate the coupling between the magnetic layers of a Co/MgO/Py MTJ. 
Two magnetic resonance peaks were observed for both magnetic layers, as probed at the Co and Ni $L_3$ X-ray absorption edges.

\begin{figure}[t]				               
\includegraphics[width = 80mm]{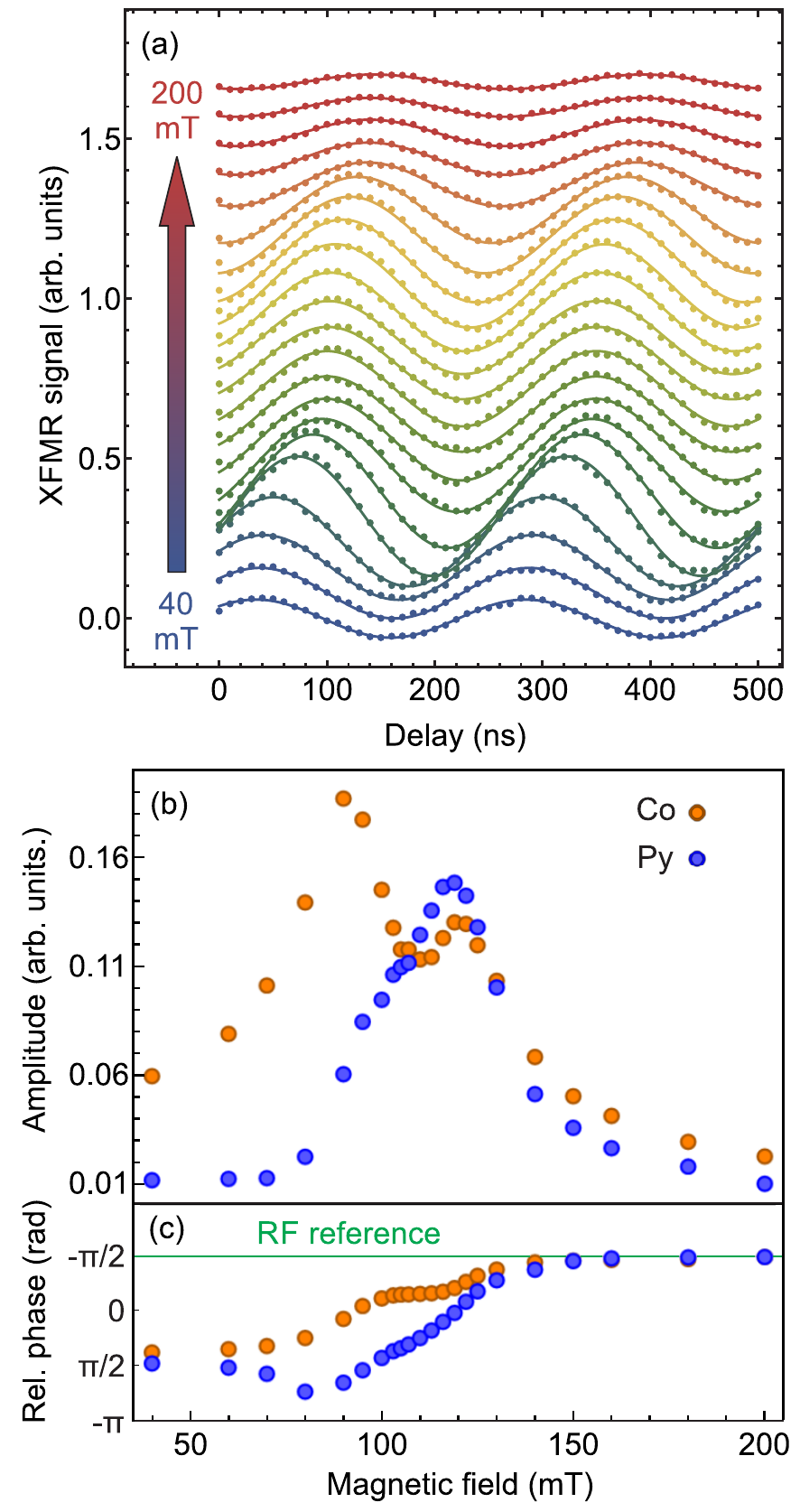}
\centering
\caption{Time resolved precession. 
(a) Series of XFMR delay scans for the Co layer in a Co/MgO/Py MTJ at 80 K continuously driven at 4 GHz.
As expected, the period of the precession is 250 ps, and the delay scan covers two periods.
 For clarity, the data points (circles) obtained at different magnetic field values (between 40 and 200 mT) are shifted by a constant offset and have been differently colored. The drawn lines
represent the fitted sinusoidal functions. Their amplitude and phase as a function of magnetic field strength is plotted in panels (b) and (c), respectively, for both the Co (orange) and Py (blue) layers.
(Adapted from G{\l}adczuk {\it{et al.}} \cite{Gladczuk2021-PRB}).
}
\label{fig:Gladczuk}
 \end{figure}

Figure \ref{fig:Gladczuk} shows XFMR delay scans for the Co layer in the Co/MgO/Py MTJ at 80 K continuously driven at 4 GHz.
The curves in Fig.~\ref{fig:Gladczuk}(a) show  a strong increase in amplitude as well as a large phase shift across the resonance at $\sim$90 mT. 
The amplitude and phase of the precession, which are extracted using Eq.~(\ref{eq:A-psi}), are shown in Fig.~\ref{fig:Gladczuk}(b) and (c), respectively, for both Co (orange) and Ni (blue) as a function of the bias field.  The amplitude curves show that the Ni resonance originating from the Py  layer around $\sim$120 mT is strongly coupled with the Co layer. On the other hand, the Co resonance around $\sim$90 mT is only weakly present in the Py layer. 
Instead of plotting amplitude $A$ and phase $\psi$, one can also plot the FMR signal in the ($X$,$Y$)-plane as a function of field \cite{Gladczuk2021-PRB}.
Since the sine and cosine functions in Eq.~(\ref{eq:sinusoidal}) are orthogonal, the estimators of $X$ and $Y$ are given by projections to
orthogonal subspaces.

A theoretical model based on the Landau\hyp{}Lifshitz\hyp{}Gilbert\hyp{}Slonczewski equation (Eq.~(\ref{eq:LLGS})) was developed, including exchange coupling and spin pumping between the magnetic layers. Fits to the experimental data were carried out, both with and without a spin pumping term, and the goodness of the fit was compared using a likelihood ratio test. This rigorous statistical approach provided an unambiguous proof of the existence of interlayer coupling mediated by spin pumping through MgO \cite{Gladczuk2021-PRB}.
It was also found that spin pumping is more effective at lower temperatures, which
agrees with the theoretical understanding. 


\subsection{XFMR of coherent spin currents in antiferromagnetic NiO}

Antiferromagnets have recently gained large interest in the field of spintronics, as they allow for faster and more robust memory operation than present technologies and as they can carry spin current over long distances. However, many fundamental physics questions about these materials regarding their spin transport properties still remain unanswered \cite{Reichlova}. 
A spin current generated by spin pumping should have a single wave mode, carrying the coherent magnetization excitation. In contrast, spin currents generated by thermal gradients produce incoherent currents with a continuum of spin excitation modes.  The magnetic excitations in antiferromagnets typically have THz frequencies, while the resonant excitation of the ferromagnetic injector is in the GHz range.  Conventional spin pumping experiments measure only the time-averaged DC component of the spin current, i.e., they cannot distinguish between GHz and THz frequencies, which is needed to determine how the spin current propagates. Alternative techniques such as XFMR are needed to measure the time-varying AC spin current. 
 
Dabrowski {\it{et al.}}\ \cite{Dabrowski2020-PRL} used XFMR to  study the coherent spin current propagation in a device with three layers, where the top (injector) and bottom (sink) layers were ferromagnetic NiFe and FeCo, respectively, and the middle layer was epitaxial NiO (001). The phase and amplitude of the magnetization precession within adjoining source and sink FM layers were detected, from which the injection and transmission of pure AC spin current through NiO can be inferred. It was found that magnetization modes in the FM layers oscillate in phase. Furthermore, the efficiency of the spin transfer varied with the thickness of the antiferromagnet, with a maximal efficiency for a 2-nm-thick layer. These results indicate that a spin current propagates coherently through the antiferromagnetic NiO layer. The AC spin current is enhanced for NiO thicknesses of less than 6 nm, both with and without a nonmagnetic spacer layer inserted into the stack, in a manner consistent with previously reported experimental measurements of DC spin current and theoretical studies \cite{Khymyn2016}. The XFMR results show that the propagation of spin current through NiO layers is mediated by evanescent antiferromagnetic spin wave modes at GHz frequencies, rather than THz frequency magnons.

 \subsection{DFMR for mode-resolved detection of magnetization dynamics}
 



\begin{figure}[t]				               
\includegraphics[width = 86mm]{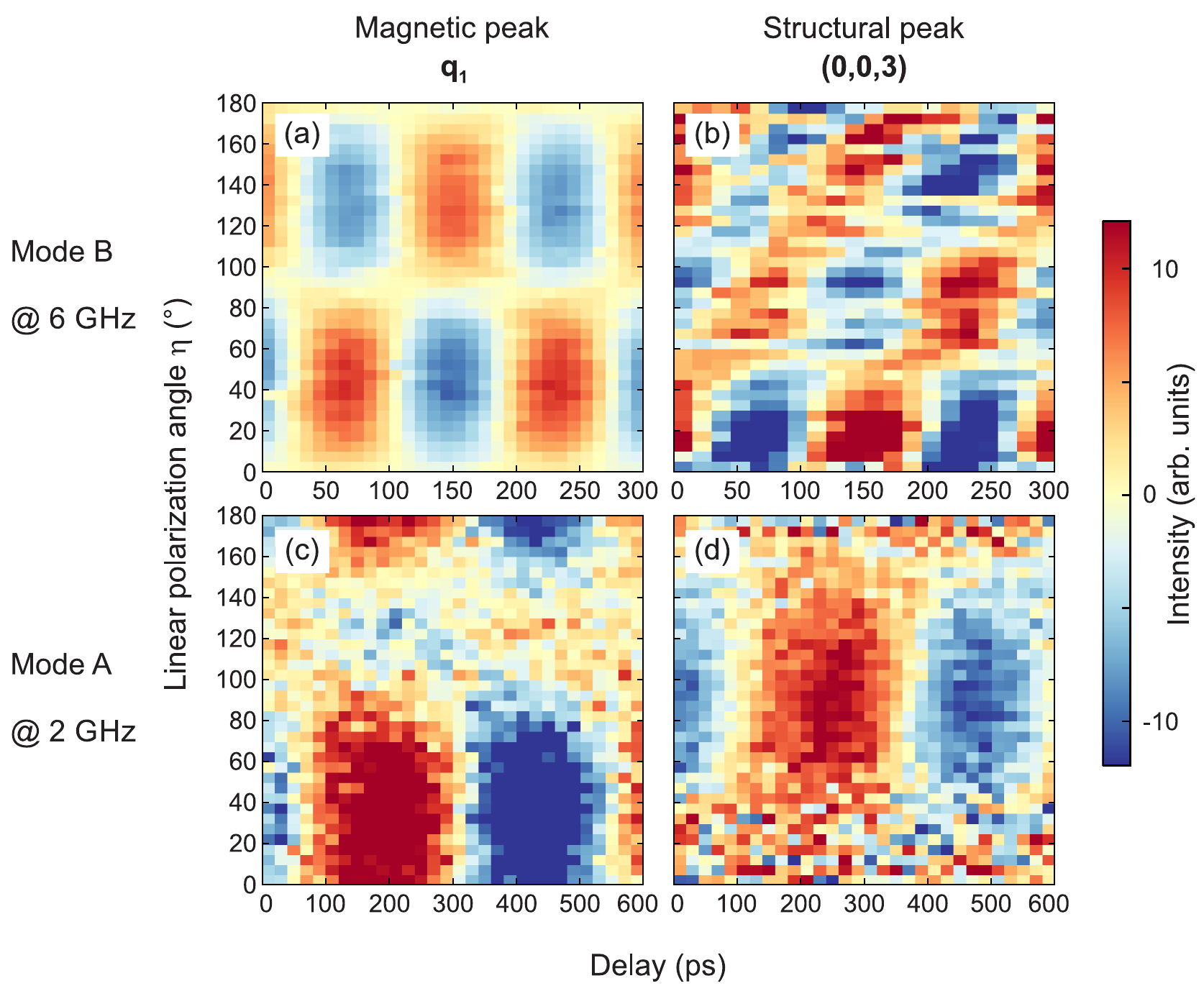}
\caption{
DFMR delay scans of the structural and magnetic peaks as a function of linear polarization angle. Measurements of (a,b) the anisotropic mode B at 6 GHz and (c,d) the isotropic mode A at 2 GHz.  
The results for the magnetic peak and the structural (0,0,3) peak are shown in   the left and   the right column, respectively. 
The magnetic resonance modes are probed with linearly polarized light for the range of incident polarization angles $\eta$ between $0$--$180^\circ$.
(Adapted from Ref.~\cite{Burn2020-DFMR}).
}
\label{fig:DFMR}
 \end{figure} 		               

Recent scientific interest has shifted towards more complex magnetically ordered materials, which are promising for high-density and low-energy consumption devices. These systems contain chiral magnetic phases such as helical, conical, or skyrmion spin structures, originating from the Dzyaloshinskii\hyp{}Moriya interaction (DMI) found in noncentrosymmetric bulk materials, as well as in systems where symmetry breaking occurs at a ferromagnetic/heavy metal interface. Such spin structures are much more complex than simple ferromagnetic structures, especially their dynamic behavior is so far ill-understood.

The periodic structure of  magnetically ordered systems can be probed by  resonant elastic X-ray scattering (REXS), making use of interference effects from the regularly repeating magnetization density variations. This  leads to pure magnetic X-ray scattering peaks which give information about the static magnetic structure. Analysis of these magnetic peaks  in REXS measurements using synchrotron radiation has led to significant progress in the understanding of chiral magnetic systems \cite{vanderLaanCRP,Zhang2016}.

In a pioneering DFMR experiment, Burn {\it{et al.}}~\cite{Burn2020-DFMR} investigated the complex dynamic behavior of the chiral spin structure in Y-type hexaferrite Ba$_2$Mg$_2$Fe$_{12}$O$_{22}$.  
 VNA-FMR measurements  of this material showed a field\hyp{}frequency map  containing two  ferromagnetic resonance modes. While mode A is isotropic, i.e., its field value is independent of the direction of the applied field, mode B is anisotropic, showing greater absorption at increasingly higher fields as the field direction rotates out-of-plane.

REXS at the Fe $L_{2,3}$ absorption edge was used to characterize
the static magnetic structure of the hexaferrite and to determine
its field dependence. Static REXS measurements along (0,0,$\ell$)
in zero field show a (0,0,3) structural peak
decorated with two incommensurate magnetic satellites.

The DFMR signal was measured by pointing the photodiode  at the scattered beam, selecting either the structural or the magnetic satellite peak (Fig.~\ref{fig:setup}).
Delay scans were measured as a function of applied field using linearly polarized X-rays.
Sinusoidal fits to the measured data enables the extraction of amplitude and phase. 
Fig.~\ref{fig:DFMR} shows the delay scans of the structural and magnetic peaks of the Y-type hexaferrite for variable incident linear polarization angles  $\eta$. 
The panels (a,b) in the top row refer to the anisotropic mode B at 6 GHz, and the panels (c,d) in the bottom to the isotropic mode A at 2 GHz. 
The left and right column refer to the results for the magnetic peak and   the structural (0,0,3) peak, respectively.  
The  results were compared to computer simulations of the Y-type hexaferrite to obtain  insight in the periodic spin structure of this material.
 %
%

 A second  example  of the use of DFMR for mode resolved detection concerns the dynamic behavior of topological spin textures and chiral magnets, which is an area of significant interest and key to the development of fast and efficient spintronics devices. 
 DFMR measurements by Burn {\it{et al.}}\ \cite{Burn2022} revealed how the time-dependence of the magnetization dynamics relate to the complex spin texture in the well-known chiral magnetic system Cu$_2$OSeO$_3$. Using  polarized soft X-rays, the dynamic excitations in all three dimensions were probed, which revealed phase shifts that were previously undetectable and indistinguishable using conventional FMR.

\begin{figure}[t]				               
\centering
\includegraphics[width = 70mm]{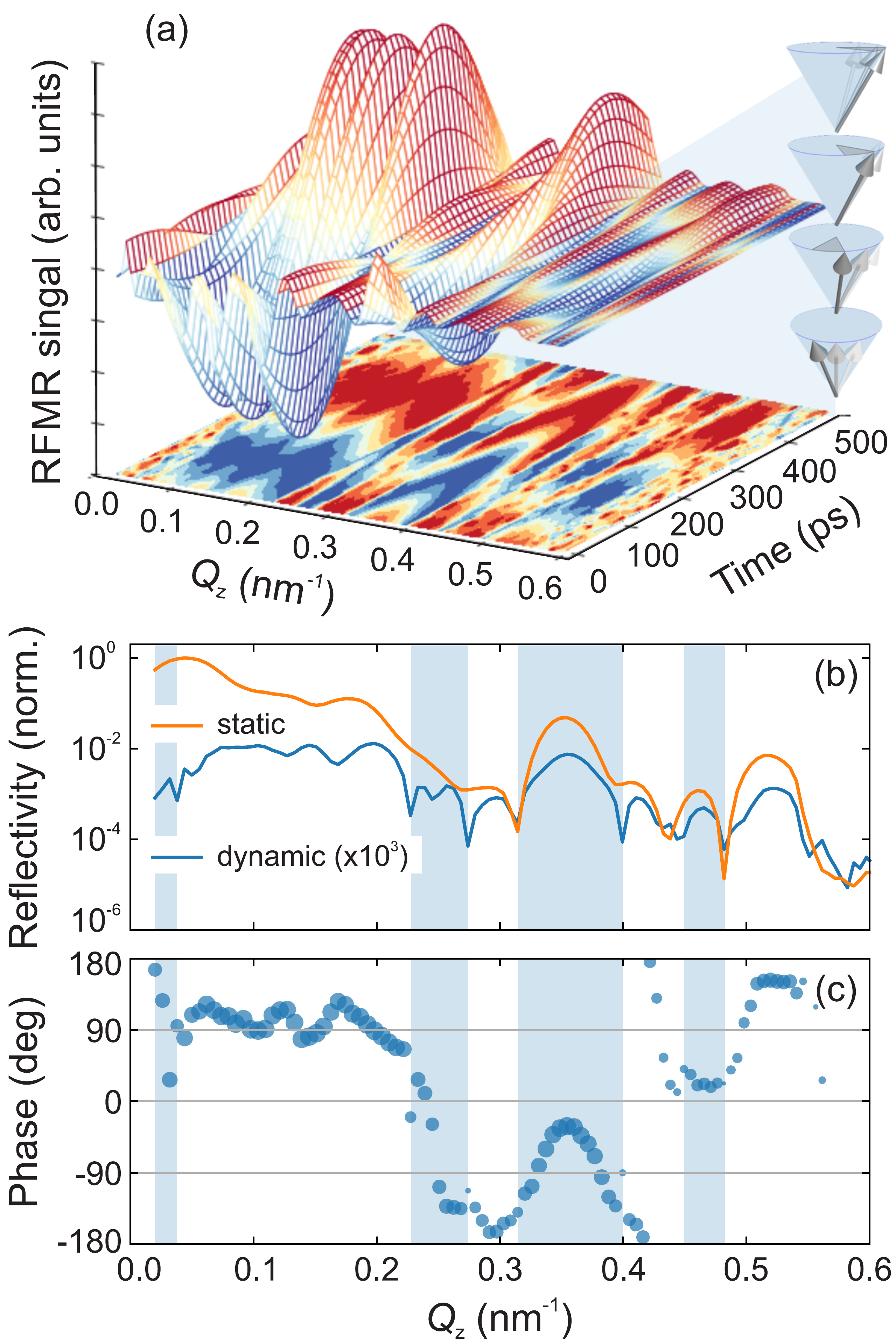}
\caption{(a)
Pseudo-3D plot of the RFMR signal and its projection showing the dynamic contribution to the
 reflectivity for a [CoFeB/MgO/Ta]$_4$ multilayer as a function of pump-probe time delay. 
The measurements were carried out with
left-circularly polarized X-rays at the  Fe $L_3$ resonance (707.7 eV) and
in an out-of-plane field of 29 mT using RF excitation
at 2 GHz. 
The various delay curves are shown for different $Q_z$, ranging between 0 and 0.6 nm$^{-1}$. The color scale represents the normalized intensity 
for each delay scan, highlighting the sinusoidal dependence and
the shift in phase as $Q_z$ is varied when the intensity is small. 
%
(b) Static and dynamic reflectivity, and (c) phase of the
dynamic reflectivity as a function of $Q_z$.  The phase point size is scaled by the strength of the dynamic signal amplitude, and the blue-shaded regions indicate where the 180$^\circ$ phase shifts have been subtracted to reveal the otherwise smooth phase variation.
(Adapted from Ref.~\cite{Burn2020-RFMR}).
}
\label{fig:RFMR}
 \end{figure} 		               

\subsection{RFMR on a [CoFeB/MgO/Ta]$_4$ multilayer}
\label{sec:RFMR}

X-ray reflectivity with the photon energy tuned to the absorption edge has become a valuable tool for characterizing the depth-dependent structure of layered materials. The  X-ray reflectivity is measured as a function of the scattering vector $Q_z$ = ${\mathbf{k}}_s - {\mathbf{k}}_i$ = $(4 \pi / \lambda)  \sin \vartheta$,  where ${\mathbf{k}}_i$  (${\mathbf{k}}_s$) is the ingoing (outgoing) wavevector of the X-rays with incident angle $\vartheta$ and  wavelength $\lambda$.
The scattering length density, which gives the scattering strength of the chemical and magnetic species within the depth profile of the film, is obtained through fitting the reflectivity data. 

Burn {\it {et al.}}~\cite{Burn2020-RFMR} revealed the depth dependence of the magnetization dynamics in a [CoFeB/MgO/Ta]$_4$ multilayer system. The structural depth profile was characterized through static X-ray reflectometry.
The dynamic reflectivity was probed with stroboscopic DFMR using an out-of-plane saturating field of $H_{\mathrm{Bias}}$ = 29 mT and an RF field generated by the CPW beneath the sample.  The RF field was phase-locked to the fourth harmonic of the $\sim$500 MHz synchrotron master clock at 2 GHz. The time dependence of the reflectivity during precession was mapped out as a function of the time delay between the RF pump and X-ray probe. Fig.~\ref{fig:RFMR}(a) shows a color map of the sinusoidal variation in the reflected signal with a 500 ps period, corresponding to the 2 GHz excitation.  The amplitude and phase of the dynamic signal are extracted by fitting the sinusoidal delay scans. The amplitude is plotted in Fig.~\ref{fig:RFMR}(b) alongside the static reflectivity for the different values of $Q_z$, ranging between 0 and 0.6 nm$^{-1}$, and the phase in Fig.~\ref{fig:RFMR}(c).

Both the static intensity and the amplitude of the dynamic signal in Fig.~\ref{fig:RFMR}(b) show reflectivity fringes resulting from interference effects arising from the layered chemical and magnetic structure. Additional minima are observed in the dynamic case. 
The phase of the dynamic signal in Fig.~\ref{fig:RFMR}(c) shows variations with two contributions. Firstly, abrupt 180$^\circ$ phase jumps occur, coinciding with minima in the amplitude of the dynamic signal. These 180$^\circ$  jumps correspond to inversion of the sign of the XMCD signal measured at different scattering conditions.
 In addition, there are smoother variations in the phase, which can be attributed to variations in the magnetization dynamics occurring between the magnetic layers in the multilayered structure.

To reveal the depth dependent magnetization dynamics, the experimental results were compared with modeling of the dynamic behavior In all layers, the magnetization precesses about a nominal static state when excited by an RF field. It was shown that inclusion of a small, but significant phase lag of 5$^\circ$  between the four layers is necessary to explain the observed change in phase of the dynamic signal. In contrast, a single slab of magnetic thin film material shows a coherent precession of the magnetization as a function of depth. 

With RFMR, the dynamics from different layers containing the same element can be explored, and this technique has the potential to study the dynamics of interfacial layers and proximity effects in complex thin film and multilayer materials for future magnetic memory and processing device applications.

\section{Conclusions}
\label{sec:Conclusions}

Although conventional FMR is a  powerful technique to study magnetic resonances in thin films and multilayers, the measured  response corresponds to an average over the entire magnetic structure of the sample.
In contrast, X-ray based FMR techniques  allow for time-resolved measurements of the magnetization dynamics,
and, in addition, offer the benefits of XMCD, such as element-, site-, and shell\hyp{}specificity \cite{GVDLAFG2014}. 
The time resolution is achieved by stroboscopic probing using higher harmonics (1-10 GHz) of the synchrotron master clock.

XFMR  can be used to study spin-transfer torque, dipolar field strength, magneto-crystalline anisotropy, interlayer exchange coupling, gyromagnetic ratio and damping constants.  
It can be applied to study the behavior of spintronics systems, e.g., spin pumping in magnetic multilayers, heterostructures, spin valves, MTJ, etc.
The amplitude and phase of the magnetic resonances extracted from the field-dependence of the precessional plots  enable us to distinguish between static and dynamic exchange coupling and to quantify their relative contributions.
Apart from measuring the signal in absorption, XFMR can also be detected in diffraction and reflectivity; each of these techniques bringing  unique advantages.
DFMR reveals the dynamical spin modes at the probed magnetic wavevectors, and RFMR gives the depth-resolved dynamics in magnetic multilayers.
Future XFMR studies can be envisaged to investigate vortex dynamics, spatial resolution imaging, and X-ray holography.

\section{Acknowledgments}
The XFMR experiments were carried out on beamline I10 at the Diamond Light Source (Oxfordshire, United Kingdom).
We like to acknowledge valuable collaborations with 
Alex A.\ Baker, 
David M.\ Burn,
Maciej Dabrowski, 
Adriana I.\ Figueroa, 
Lukasz Gladczuk, 
and Robert J.\ Hicken.

--------------------------

 \bibliographystyle{model1a-num-names}

\end{document}